\documentclass[journal]{IEEEtran}
\usepackage{subcaption}
\usepackage{cite}
\usepackage{amsthm}
\theoremstyle{definition}
\newtheorem{definition}{Definition}[section]
\usepackage{stfloats}
\usepackage{amsmath}
\usepackage{hyperref}
\usepackage{cleveref}
\usepackage{algorithm}
\usepackage{algorithmic}
\usepackage{pifont}
\newcommand{\cmark}{\ding{51}}  
\newcommand{\xmark}{\ding{55}}  
\usepackage{booktabs}

\usepackage{multirow}
\ifCLASSINFOpdf
  \usepackage[pdftex]{graphicx}

\else

\fi

\usepackage{algorithmic}

\usepackage{array}

\usepackage{fixltx2e}

\usepackage{url}

\title{ELEAT-SAGA: Early \& Late Integration with Evading Alternating Training for Spoof-Robust Speaker Verification}
\author{Amro~Asali$^{1}$, Yehuda~Ben-Shimol$^{1}$, and Itshak~Lapidot$^{2,3}$%
\thanks{$^{1}$School of Electrical and Computer Engineering, Ben Gurion University of the Negev, Israel}%
\thanks{$^{2}$School of Electrical Engineering, Afeka Academic College of Engineering in Tel Aviv, Israel}%
\thanks{$^{3}$LIA, Avignon University, France}%
\thanks{Contact e-mails: asaliam@bgu.ac.il, benshimo@bgu.ac.il, itshakl@afeka.ac.il}%
}
\begin{document}
\maketitle

\begin{abstract}
\textit{Spoofing-robust automatic speaker verification} (SASV) seeks to build automatic speaker verification systems that are robust against both zero-effort impostor attacks and sophisticated spoofing techniques such as \textit{voice conversion} (VC) and \textit{text-to-speech} (TTS). In this work, we propose a novel SASV architecture that introduces \textit{score-aware gated attention (SAGA)}, SASV-SAGA,  enabling dynamic modulation of speaker embeddings based on \textit{countermeasure} (CM) scores. By integrating speaker embeddings and CM scores from pre-trained ECAPA-TDNN and AASIST models respectively, we explore several integration strategies including early, late, and full integration. We further introduce \textit{alternating training for multi-module} (ATMM) and a refined variant, \textit{evading alternating training} (EAT). Experimental results on the ASVspoof 2019 Logical Access (LA) and Spoofceleb datasets demonstrate significant improvements over baselines, achieving a \textit{spoofing-aware speaker verification equal error rate}  (SASV-EER) of 1.22\% and \textit{minimum normalized agnostic detection cost function} (min a-DCF) of 0.0304 on the ASVspoof 2019 evaluation set. These results confirm the effectiveness of score-aware attention mechanisms and alternating training strategies in enhancing the robustness of SASV systems.
\end{abstract}


\IEEEpeerreviewmaketitle

\section{Introduction}
\IEEEPARstart{A}{\MakeLowercase{\textit{utomatic}}} \textit{speaker verification} (ASV) systems offer a convenient approach to biometric authentication, wherein the system must determine whether a test utterance was spoken by a claimed target speaker~\cite{wu2015spoofing}. Most of the current ASV systems are typically based on \textit{deep neural networks} (DNNs) and perform effectively in distinguishing target utterances from zero-effort non-targets~\cite{snyder2017deep,Desplanques-2020,ggemini}. However, their performance deteriorates when confronted with manipulated or synthesized (i.e., spoofed) speech~\cite{evans2013spoofing,wu2015spoofing}. Recent studies have shown that ASV systems are evolving to effectively reject spoofed inputs, even in zero-shot scenarios~\cite{jung2024extentasvsystemsnaturally}. However, rapid advances in speech manipulation techniques, such as \textit{text-to-speech} (TTS) and \textit{voice conversion} (VC), highlight the continued need to improve SASV systems. Modern SASV systems exhibit superior performance compared to conventional ASV systems, as reflected by lower SASV-EER scores across target, zero effort non-target, and spoofed non-target trials~\cite{jung2022sasv2022spoofingawarespeaker,Liu-2024,probfusi}.
Reliable speaker verification typically comprises two subsystems: ASV~\cite{ggemini,Desplanques-2020} and spoofing CM classifiers~\cite{jung2021aasistaudioantispoofingusing,tak2021end,borodin24_asvspoof}. A key research question is how to best integrate these subsystems to ensure robust verification. Common approaches to integrating the systems include cascaded (tandem)~\cite{sahidullah2016integrated,shim22_odyssey,wu2015spoofing,avishai,weizman2024tandemspoofingrobustautomaticspeaker} and in-parallel (fusion) configurations, with fusion typically executed at the score or embedding levels~\cite{shim22_odyssey,sahidullah2016integrated,probfusi,Liu-2024}. While the results reported in  in \cite{vicomtech22_iberspeech} indicate that integrated systems perform significantly better on the joint task, they often exhibit notable performance degradation compared to standalone systems when evaluated on each task separately. Tak et al. ~\cite{ge2022potentialjointlyoptimised} found that joint optimization, whereby the ASV and CM systems utilized for feature extraction are updated during the SASV training process, can enhance reliability at the SASV level. However, integrating fixed, pre-trained subsystems often results in superior performance.

The findings in~\cite{shim22_odyssey,jung2022sasv2022spoofingawarespeaker} indicate that simple backend fusion methods that utilize distinct using CM and ASV scores may be a viable short-term solution. Nevertheless, the authors suggest, and the results demonstrate that more sophisticated strategies, particularly those employing DNN integrations or fully integrated systems, offer greater potential by effectively exploiting the synergy between the CM and ASV components.

In this work we propose a novel integration architecture and training procedure aimed at enhancing both performance and generalization to unseen spoofing attacks. The experimental evaluation is conducted on the ASVspoof2019 Logical Access dataset, which serves as a standardized benchmark~\cite{wang2020asvspoof2019largescalepublic}.  Additionally, experiments are conducted on the SpoofCeleb dataset, which has recently attracted increased attention in the SASV community as an in-the-wild SASV benchmark~\cite{spoofceleb}. A preliminary version of our work was presented in~\cite{asali25_interspeech}.

The remainder of this paper is organized as follows. \autoref{sec:background} reviews the background and related work. \autoref{sec:efusion} and \autoref{sec:SAGA_system} detail our proposed SAGA architecture and novel training procedures. The subsequent sections present our experimental setup, results, and analysis. Finally, section \autoref{sec:conclusion} concludes the paper.

\section{Background}
\label{sec:background}
This section presents a concise overview of the ASV and CM systems employed for embedding extraction, along with a review of existing literature on proposed solutions to the SASV problem.
\subsection{Literature review of SASV solutions}
This subsection presents a brief overview of key systems proposed to solve the SASV problem, defined as follows: 
\begin{definition}[SASV] 
Given a pair: enrollment utterances $\left\{U_{\mathrm{erl}}\right\}$ of the target speaker and a test utterance $U_{\mathrm{tst}}$, the SASV system evaluates whether $U_{\mathrm{tst}}$ was spoken by the target speaker (output $y^\mathrm{SASV}=1)$ or non-target ($y^\mathrm{SASV}=0$). Non-target can be a zero-effort impostor attack or a spoofing attack.
\end{definition}

The ASVspoof challenge has been shown to extend the findings of established research in anti-spoofing, albeit from the perspective of treating it as a discrete task rather than as an integrated component of ASV \cite{wang2020asvspoof2019largescalepublic}. To address this separation, the SASV challenge was introduced to promote integrated, jointly optimized systems \cite{jung2022sasv2022spoofingawarespeaker}. This new challenge employs the ASVspoof 2019 Logical Access dataset for its unified evaluation protocol.

The baseline systems developed for the SASV2022 challenge~\cite{shim22_odyssey,jung2022sasv2022spoofingawarespeaker} introduce two main fusion strategies. The first baseline performs score fusion through summation, after applying a sigmoid function to the CM output to constrain it within the [0,1] range before fusing with the ASV cosine similarity score between enrollment and test embeddings reaching 1.71\% SASV-EER on the evaluation set.
The second baseline is a DNN-based fusion strategy that combines speaker embeddings  (ASV) from enrollment and test utterances, with a CM embedding from the test utterance using a three-layer fully connected network with leaky ReLU activations. This model classifies inputs as either target or non-target, achieving an SASV-EER of 6.73\%.

In~\cite{probfusi}, the authors proposed a probabilistic framework for fusing scores from ASV and CM subsystems. They further introduced strategies for direct inference and fine-tuning to predict SASV scores, achieving an SASV-EER of 1.53\%.

In~\cite{vicomtech22_iberspeech}, a shallow neural network processes the test utterance’s verification and spoofing embeddings to compute a spoof-based score, which is then combined with the cosine similarity between the speaker embeddings to make the final decision. This method achieved an SASV-EER of 0.84\%. They also observed that using both ASV and CM test embeddings for the spoofing score improved performance, suggesting that speaker embeddings may contain discriminative anti-spoofing features.

An alternative approach, the generalized spoof-aware ASV (G-SASV) system, integrates limited CM training data into a unified DNN-based backend through domain adaptation and multi-task learning ~\cite{Liu-2024}. This system eliminates the need for a separate CM module (CM embeddings) during authentication and achieved an SASV-EER of 8.62\%, indicating that while this is a novel and promising approach, more work is required to compensate for the performance degradation caused by removing the CM from the authentication stage.

The authors in~\cite{shim22_odyssey,jung2022sasv2022spoofingawarespeaker} also reported on a cascaded combination of CM and ASV systems. This approach is not considered an official SASV baseline, since fusion is performed at the decision level and only trial-level outputs are available. Consequently, trials rejected by the CM lack corresponding ASV scores, making EER estimation unfeasible. Instead, the method is evaluated using the Half Total Error Rate (HTER), for which a value of 1.47\% was reported. The authors cautioned that lower HTER scores in cascaded systems should not be interpreted as superior overall performance, emphasizing that EER and HTER results are not directly comparable.

\subsection{Automatic speaker verification system}
In this work, we employ the \textit{emphasized channel attention, propagation, and aggregation time delay neural network} (ECAPA-TDNN) for speaker verification~\cite{Desplanques-2020}. The system operates on 80-dimensional \textit{Mel-frequency cepstral coefficients} (MFCCs) and uses a modified Res2Net architecture as its backbone processing block, together with dimensional \textit{squeeze-excitation} (SE) blocks to model global channel interdependencies. The network further incorporates attentive statistics pooling, which allows it to focus on informative speech frames, and multi-layer feature aggregation that combines both shallow and high-level speaker characteristics into utterance-level representations. Speaker embeddings are extracted from the final hidden layer and possess a dimensionality of 192.

\subsection{CM system} 
We employ the \textit{audio anti-spoofing using integrated spectro temporal graph attention networks} (AASIST) framework, which combines a RawNet2 front end with a graph attention architecture~\cite{jung2021aasistaudioantispoofingusing}. The model uses a Sinc based convolutional encoder to derive time frequency representations, which are subsequently processed by a residual network to capture higher level features. Two graph attention modules are then applied to model spectral and temporal relationships, respectively. The countermeasure embeddings yielded by the system possess a dimensionality of 160. 

\section{Proposed Embedding Fusion Baseline \label{sec:efusion}}
\begin{figure}
    \centering
    \includegraphics[width=0.8\linewidth]{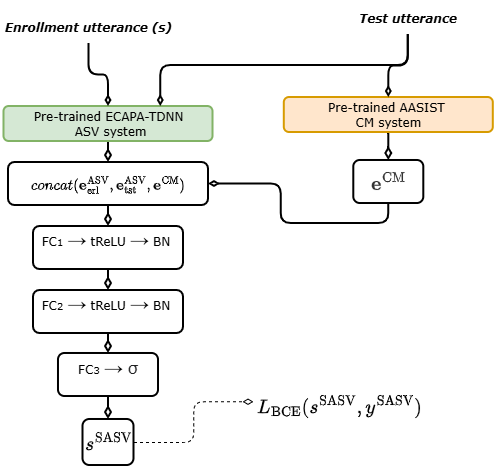}
    \caption{Diagram of the eFusion SASV system. Dashed arrows denote operations exclusive to training stage.}
    \label{fig:efusion}
\end{figure}
In this study, we investigate the integration of speaker embeddings, derived from a pre-trained ECAPA-TDNN model \cite{Desplanques-2020}, with CM embeddings, derived from a pre-trained AASIST \cite{jung2021aasistaudioantispoofingusing}. We begin with the classic embedding fusion and systematically modify \lq Baseline 2\rq\   from \cite{jung2022sasv2022spoofingawarespeaker} to improve its efficacy. The resulting model ultimately serves as the primary baseline for our proposed framework, which we refer to as \textit{embedding fusion} (eFusion) illustrated in \autoref{fig:efusion}. The first set of changes made to this system is replacing the LeakyReLU activation functions with parameter shared tReLU activation functions, following a formulation similar to that proposed in~\cite{Liu-2024}. This modification was found to consistently outperform LeakyReLU in our preliminary experiments. Consider a linear transformation followed by a nonlinearity. Instead of applying a fixed activation, we introduce a learnable transformation prior to rectification. Let the affine mapping be parameterized by weights $W_i$ and bias $b_i$. We define a transformed rectified linear unit that applies an additional trainable mapping before thresholding:
\begin{equation}
t\mathrm{ReLU}_{W_a}(W_i x + b_i) =
\max\!\left(W_a (W_i x + b_i), \mathbf{0}\right)
\label{equation:frelu}
\end{equation}
where $W_a$ is a learnable matrix initialized as the identity matrix. The rectification is applied element-wise after the transformation. This formulation generalizes the standard ReLU by allowing the activation geometry to be adapted during training. Related formulations with constrained parameterization have been explored in prior work~\cite{pact,Bell-2021,Liu-2024}. We have also introduced some regularization such as \textit{batch normalization} (BN) layers and weight decay.

\section{SAGA: Score Aware Gated Attention}
\label{sec:SAGA_system}
\begin{figure}[t]
   \centering
   \includegraphics[width=0.8\linewidth]{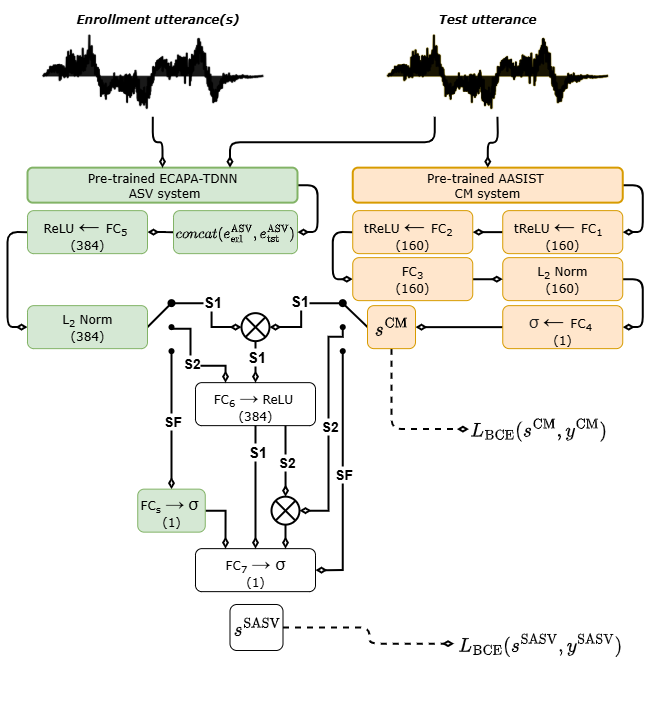}
   \caption{Diagram of the proposed SAGA system, illustrating various CM score integration strategies. Dashed arrows denote operations exclusive to the training stage.}
   \label{fig:atmmsaga}
\end{figure}

The findings in ~\cite{vicomtech22_iberspeech} suggest that acquiring correlation information from the three distinct embeddings (i.e., ASV Enroll, ASV Test, CM Test embeddings) poses a considerable challenge. Consequently, exploring the integration of speaker embeddings with a CM score emerges as a viable approach.
While this integration framework is general and can accommodate different ASV and CM backbones, in our ASVspoof2019 experiments the speaker embeddings are derived from a pre-trained ECAPA-TDNN model. For the SpoofCeleb experiments, in addition to ECAPA-TDNN, we also conduct experiments in which ECAPA-TDNN is replaced with SKA-TDNN~\cite{mun2023frequency}, while the CM score is computed using embeddings extracted by a pre-trained AASIST model. The CM score is applied as a multiplicative gate on the normalized speaker embeddings, producing spoofing-aware representations. We denote this mechanism as SAGA. These embeddings are then utilized to calculate SASV scores. We investigate two integration levels while preserving the overall system architecture. We also explore a score fusion strategy for a more comprehensive comparative analysis. The proposed system is illustrated in~\autoref{fig:atmmsaga}. 

\subsubsection{S1: Early Integration}
The S1 strategy follows the path labeled ``S1" in~\autoref{fig:atmmsaga}. In this configuration, the CM embeddings extracted by AASIST are forwarded through two fully connected layers along the upper-right branch of the architecture (shown in orange in the figure). These representations are transformed using parameter-shared $t\mathrm{ReLU}$ activations, as defined in \autoref{equation:frelu}. The resulting embeddings are then processed by an additional fully connected layer and normalized via L2-normalization. Finally, a fully connected layer with a Sigmoid activation produces the CM score, denoted as $\mathrm{s^{CM} \in [0,1]}$.

In parallel, on the left processing path (colored green in ~\autoref{fig:atmmsaga}), the ECAPA-TDNN speaker embeddings are concatenated and passed through a fully connected layer with a standard ReLU activation. The resulting representations are L2-normalized ensuring a consistent embedding scale prior to modulation by the CM score, yielding ${\mathbf{e}^{\mathrm{ASV}}}$.
To incorporate SAGA, we apply a multiplicative gating mechanism in which the CM score acts as an adaptive gate over the normalized speaker embeddings:
\begin{equation}
\label{equation:saga_equation}
\mathbf{e}^\mathrm{SASV} = g(s^\mathrm{CM}, \mathbf{e}^\mathrm{ASV}) = s^\mathrm{CM}\mathbf{e}^\mathrm{ASV}
\end{equation} where
 ${g(s^\mathrm{CM}, \mathbf{e}^\mathrm{ASV})}$ represents the SAGA operation. As a result, spoofed trials (${s^{\mathrm{CM}}\approx0}$) are suppressed, whereas bona fide trials (${s^{\mathrm{CM}}\approx1}$) remain largely unaffected. $\mathbf{e}^\mathrm{SASV}$ represents the spoofing-robust speaker verification embeddings. The resulting embeddings are further transformed by a fully connected layer with ReLU activation, followed by a final fully connected layer with a Sigmoid activation that produces the SASV score.
\subsubsection{S2: Late Integration}
Although this strategy shares the same overall architecture as S1, the SAGA operation is applied at a later stage in the ASV processing path (denoted as ``S2" in~\autoref{fig:atmmsaga}).
\subsubsection{S3: Full Integration}
A full (i.e., both early and late) integration strategy combines both S1 and S2 approaches, applying the SAGA operation twice to enhance performance by reinforcing spoof awareness at multiple stages of the system’s processing procedure.
\subsubsection{SF: Score Fusion}  
This strategy performs score-level fusion of the ASV and CM outputs using a fully connected layer, yielding the SASV score. The resulting system (labeled ``SF" in~\autoref{fig:atmmsaga}) serves as a comparative baseline.
\subsection{Loss function}
  To optimize both the speaker verification and countermeasure objectives, a multi-task learning framework is employed, following the approach described in~\cite{Li2020JointDecision}.

\begin{equation}
    L^{\mathrm{total}} = \lambda \cdot L^{\mathrm{SASV}}_{\mathrm{BCE}}(s^\mathrm{SASV},y^\mathrm{SASV}) + (1 - \lambda) \cdot L^{\mathrm{CM}}_{\mathrm{BCE}}(s^\mathrm{CM},y^\mathrm{CM})
    \label{equation:total_loss}
\end{equation}
where $\lambda \in [0,1]$ controls the relative weighting of the two objectives. $L^{\mathrm{SASV}}_{\mathrm{BCE}}(\cdot,\cdot)$ is the \textit{binary cross-entropy} (BCE) loss computed between the system’s SASV output score $s^{\mathrm{SASV}}\in[0,1]$ and the corresponding label $y^{\mathrm{SASV}}\in\{0,1\}$. Similarly, $L^{\mathrm{CM}}_{\mathrm{BCE}}(\cdot,\cdot)$ denotes the BCE loss evaluated between the CM output score $s^{\mathrm{CM}}\in[0,1]$ and the CM label $y^{\mathrm{CM}}\in\{0,1\}$, where $y^{\mathrm{CM}}=1$ indicates a bona fide trial and $y^{\mathrm{CM}}=0$ a spoofed trial.

\subsection{ATMM-SAGA: alternating training for score aware gated attention}
\label{subsection:ATMM_SAGA}
The ATMM strategy is motivated by key limitations observed in traditional \textit{multi-task learning} (MTL), especially under domain mismatch. In spoofing-robust speaker verification, the ASV and CM tasks are trained on distinct distributions, often leading to gradient interference during a joint optimization. Yu et al.~\cite{yu2020gradient} showed that conflicting task gradients can harm performance and proposed \textit{project conflicting gradients} (PCGrad) to project gradients onto non-conflicting directions. Chen et al.~\cite{chen2018gradnorm} introduced \textit{gradient normalization} (GradNorm) to dynamically reweigh task losses for balanced learning. In this work, we suggest an alternating approach. Here we introduce a modified training procedure from continuous training of all the model layers to alternating training. A switching training objective focuses the weight updates on different modules of the system. This method of training also prevents overfitting and makes SASV training more efficient.

\subsubsection{Alternating Training for Multi-Module}
During the training phase, two separate training datasets are utilized, as described in~\autoref{table:training data}. The first dataset is utilized for the training of spoofing countermeasures, while the second dataset is solely employed for speaker verification training. For the CM training dataset, the ASVspoof2019 LA train set \cite{wang2020asvspoof2019largescalepublic} is employed to generate pairs of enrollment and test utterances. Specifically, each bona fide utterance is paired with a random subset of the same speaker's bona fide utterances to form target trials and with other speakers' bona fide utterances to form zero-effort non-target trials. Additionally, we have paired bona fide utterances with the same speaker's spoofed utterances to create spoof non-target trials. The speaker verification training dataset is comprised of the VoxCeleb1 E and H partitions, and the ASVspoof2019 LA training set bona fide pairs.
\begin{table}[!t]
\renewcommand{\arraystretch}{1.3}
\caption{The training datasets used in our experiments, detailing the number of target, non-target, and spoofed trials for both the spoofing CM and speaker verification datasets.}
\label{table:training data}
\centering
\begin{tabular}{|l|c|c|c|}
\hline
\textbf{Dataset} & \textbf{Target} & \textbf{Non-target} & \textbf{Spoof} \\
\hline
Spoofing CM & 262228 & 249094 & 463910 \\
\hline
Speaker verification & 806025 & 779601 & 0 \\
\hline
\end{tabular}
\end{table}

We proposed ATMM, where each task parameters are trained with an alternating scheduling on the specific task data while the other is frozen. Furthermore, to ensure that the post-gate SASV layers can learn the gate dynamics on spoofed samples, \(\lambda\) should not be exactly zero. For example, during CM-focused updates  \(\lambda = 0.1\) so that \(L_{\mathrm{SASV}}\) remains present in both \(p=0\) and \(p=1\) phases. This ensures that the introduced shared representation layers ($FC_6$ and $FC_7$ in \autoref{fig:atmmsaga}) are trained with both tasks and get a spoofing aware representation. To adapt to the gate dynamics, bona fide data is present in both $p=0$ and $p=1$ and spoofed data is present only in $p=0$. The ATMM training strategy facilitates joint optimization by alternating parameter updates across the two modalities. At each iteration, a random binary choice determines whether the update focuses on the ASV or CM component, which helps mitigate overfitting to a single task through selective backpropagation and targeted weight freezing. Notably, training for both objectives requires paired enrollment and test utterances, as the shared representation layers remain unfrozen throughout the process. A complete description of a single training round, defined as one randomly selected pass over the data pool (which may include all available data or only a subset), is presented in Algorithm~\ref{alg:atmm}.

\begin{algorithm}[t]
\caption{ATMM: training for a single round}
\label{alg:atmm}
\begin{algorithmic}[1]
\FOR{$N$ iterations}
    \STATE Choose randomly $p \in \{0, 1\}$.
    \IF{$p = 0$}
        \STATE Set $\lambda \gets 0.1$.
        \STATE Sample $1/N$ of the spoofing CM dataset.
        \STATE Freeze the speaker verification weights (colored in green), as shown in Figure~\ref{fig:atmmsaga}.  
    \ELSE
        \STATE Set $\lambda \gets 0.9$.
        \STATE Sample $1/N$ of the speaker verification dataset.
        \STATE Freeze the CM weights (colored in orange), as shown in Figure~\ref{fig:atmmsaga}.
    \ENDIF
    \STATE Compute the total loss according to \autoref{equation:total_loss}.
    \STATE Perform backpropagation and update the weights of the unfrozen components.
\ENDFOR
\end{algorithmic}
\end{algorithm}

\subsection{ELEAT-SAGA: early and late integration with evading alternating training SAGA}
\label{subsection:ELEAT}
\begin{figure}[t]
   \centering
   \includegraphics[width=0.8\linewidth]{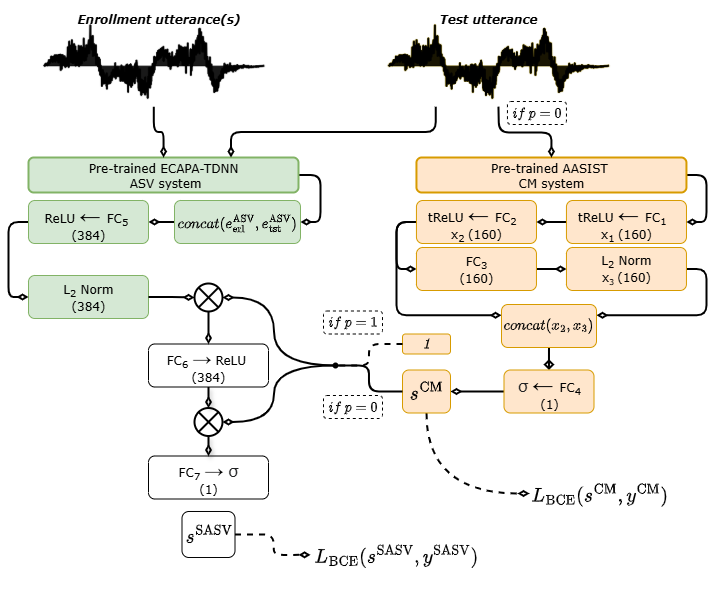}
   \caption{Diagram of the proposed ELEAT-SAGA system, illustrating branched feature extraction, and the CM bypass available only during ASV module training.}
   \label{fig:eatfi}
\end{figure}

This system is similar to ATMM-SAGA S3 (full integration), but with key modifications to improve performance. The change adds early features from the CM processing path (outlined in orange in \autoref{fig:eatfi}) to calculate the CM score. It follows a branching approach similar to that described in ~\cite{earlyexit22}. This allows early layers to capture important features that are not present in the outputs of subsequent layers. In this setup, we utilize embeddings extracted after the second activation layer, denoted as $\mathbf{x_2}$, and the $L_2$ normalization layer, $\mathbf{x_3}$, as illustrated in \autoref{fig:eatfi}.

We have also refined the ATMM algorithm by introducing EAT, as described in Algorithm \ref{alg:eat}. This modification incorporates a SAGA evading mechanism (bypass) whenever training with the speaker verification dataset (described in \autoref{table:training data}), by setting \( s^\mathrm{CM} = 1 \) in \autoref{equation:saga_equation}.
The rationale behind this modification is based on two key observations. First, the speaker verification dataset was originally constructed for training models on the ASV task. Consequently, we utilized out-of-domain data, specifically from the VoxCeleb1 dataset. Forcing the model to extract a score from CM embeddings of out-of-domain VoxCeleb1 samples would theoretically degrade performance. Second, since AASIST was not trained on VoxCeleb1, the embeddings it generates for such samples are likely to be suboptimal. Given these considerations, and the fact that the speaker verification dataset contains only bona fide utterances, the evading mechanism is a reasonable adaptation.

\begin{algorithm}[t]
\caption{EAT: training for a single round}
\label{alg:eat}
\begin{algorithmic}[1]
\FOR{$N$ iterations}
    \STATE Choose randomly $p \in \{0, 1\}$.
    \IF{$p = 0$}
        \STATE Set $\lambda \gets 0.1$.
        \STATE Sample $1/N$ of the spoofing CM dataset.
        \STATE Freeze the speaker verification weights (colored in green), as shown in Figure~\ref{fig:eatfi}.  
    \ELSE
        \STATE Set $\lambda \gets 1$.
        \STATE Set $S^{\mathrm{CM}} \gets 1.0$.
        \STATE Sample $1/N$ of the speaker verification dataset.
        \STATE Freeze the CM weights (colored in orange), as shown in Figure~\ref{fig:eatfi}.
    \ENDIF
    \STATE Compute the total loss according to \autoref{equation:total_loss}.
    \STATE Perform backpropagation and update the weights of the unfrozen components.
\ENDFOR
\end{algorithmic}
\end{algorithm}

\section{Experimental setup}
\label{sec: setup}
In the following, we describe the experimental setup for the proposed SASV system and detail the datasets used for evaluation and the metrics employed to assess performance.

\subsection{Datasets}
The availability of datasets with official SASV protocols is limited. To provide a more comprehensive evaluation of the system, it was evaluated on two datasets: ASVspoof2019 and SpoofCeleb.
\subsubsection{ASVspoof2019}
ASVspoof 2019 LA is a standard benchmark dataset for evaluating spoofing-robust speaker verification systems~\cite{wang2020asvspoof2019largescalepublic}. It consists of genuine speech and spoofed utterances produced using multiple TTS and VC techniques. The dataset is organized into Train, Dev, and Eval splits with predefined evaluation protocols, and each trial is associated with multiple enrollment utterances for target speaker modeling. While the Train and Dev partitions contain the same six spoofing attacks (A01–A06), the Eval set features 11 unseen attacks (A07–A15, A17, A18) and two additional attacks (A16, A19) derived from related synthesis algorithms.

\subsubsection{SpoofCeleb}
SpoofCeleb is a large-scale in-the-wild corpus for spoofing detection and SASV~\cite{spoofceleb}. It is derived from VoxCeleb1 through an automated pipeline that transcribes, segments, filters, and enhances speech to produce the \textit{TTS in the wild easy} (TITW-Easy) bona fide set. Using TITW-Easy, 23 TTS systems are trained to generate diverse spoofing attacks covering acoustic, vocoder, and end-to-end models. The dataset contains over 2.5M utterances from 1,251 speakers and is split into Train, Dev, and Eval partitions with non-overlapping speakers. The Train set includes 10 spoofing attacks, the Dev set includes 6 partially known or configuration-variant attacks, and the Eval set includes 9 unseen attacks. SpoofCeleb provides official spoofing detection and SASV protocols, with 55,741 utterances and 39,353 SASV trials in Dev, and 91,130 utterances and 133,448 SASV trials in Eval. Similar to ASVspoof2019, SpoofCeleb does not provide an official SASV training protocol consisting of predefined enrollment–test trials, and therefore the training protocol was constructed by us.

\subsection{Evaluation Metrics}
The performance of spoofing-robust speaker verification systems is evaluated using established metrics that quantify both discrimination accuracy and robustness in the presence of spoofing attacks. Following prior SASV work~\cite{jung2022sasv2022spoofingawarespeaker,spoofceleb}, SASV-EER and min a-DCF to provide complementary perspectives on system behavior. 

\subsubsection{EER Metrics}
EER-based evaluation treats SASV as a binary decision problem between \emph{target} and \emph{non-target} hypotheses. The non-target class comprises both zero-effort non-target (bona fide impostor) trials and spoofed non-target trials. The primary metric, referred to as SASV-EER, is defined as the error rate at the operating threshold where the false rejection rate of target trials equals the combined false acceptance rate of bona fide and spoofed non-target trials. As such, SASV-EER provides a unified measure that reflects system robustness to both speaker mismatch and spoofing attacks.



\subsubsection{min a-DCF}
While SASV-EER provides a threshold-independent measure of discrimination, it does not account for differences in class priors or the relative costs of errors. To address this limitation, we additionally report the min a-DCF, which evaluates system performance under a Bayes risk framework \cite{adcf}:
\begin{equation}
\min_\mathrm{t} \frac{ \mathrm{a\text{-}DCF}(t)}{\mathrm{{a\text{-}DCF}_{def}}}
\end{equation}
where 
\begin{align}
\text{a-DCF}(t) :=\; & C_{\text{miss}} \cdot \pi_{\text{tar}} \cdot P_{\text{miss}}(t)
                     +C_{\text{fa}}^{\text{non}} \cdot \pi_{\text{non}} \cdot P_{\text{fa}}^{\text{non}}(t) \notag \\
                     +\; & C_{\text{fa}}^{\text{spoof}} \cdot \pi_{\text{spoof}} \cdot P_{\text{fa}}^{\text{spoof}}(t)
\end{align}

\begin{equation}
\text{a-DCF}_{\text{def}} := \min \left\{ C_{\text{miss}} \pi_{\text{tar}}, C_{\text{fa,non}} \pi_{\text{non}} + C_{\text{fa,spf}} \pi_{\text{spf}} \right\}
\end{equation}
\( P_{\text{miss}}, P_{\text{fa,non}}, P_{\text{fa,spf}} \) are the miss rate (false rejection), zero-effort non-target false acceptance rate, and spoof false acceptance rate at threshold \( t \), respectively. \( \pi_{\text{tar}}, \pi_{\text{non}}, \pi_{\text{spf}} \) denote the priors for target, zero-effort non-target, and spoof trials, while \( C_{\text{miss}}, C_{\text{fa,non}}, C_{\text{fa,spf}} \) are their respective detection costs, respectively.




In contrast to EER-based metrics, min a-DCF explicitly incorporates operating costs and prior probabilities, providing a cost-sensitive evaluation that is more closely aligned with practical deployment scenarios. Together, EER-based measures and min a-DCF offer a comprehensive assessment of SASV system performance.

\section{Experiments and Results}
\label{sec: experiments and results}
This section details our experimental evaluation. We begin with experiments on the ASVspoof2019 dataset, starting from the eFusion baseline and progressively analyzing the proposed ATMM-SAGA framework, exploring various training approaches. Finally, we introduce the architectural enhancements that culminate in our ELEAT-SAGA model, concluding with a qualitative embedding analysis and a comprehensive comparison against closely related state-of-the-art systems that also rely on ECAPA-TDNN and AASIST. Finally, we extend the evaluation to the SpoofCeleb dataset to assess generalization in an in-the-wild scenario.
\subsection{Proposed Embedding Fusion Baseline}
We utilized the eFusion system described in \autoref{sec:efusion} to serve as a modified baseline for embedding fusion method in our framework, and trained it on the complete set of approximately 2 million samples, constructed from trials in the ASVspoof2019 and VoxCeleb1 datasets. Training was performed with a learning rate of \(10^{-4}\) and an $L_2$ weight decay regularization of \(10^{-7}\). To optimize performance, we experimented with different numbers of training epochs, selecting an early stopping checkpoint based on the validation loss observed on the ASVspoof2019 development dataset.

\begin{table}[!t]
\renewcommand{\arraystretch}{1.3}
\caption{Evaluation results of the eFusion system across different training epochs, compared to the Baseline2 model. Performance is measured in terms of SASV-EER (\%) and min a-DCF on the ASVspoof2019 LA dataset’s development and evaluation sets.}
\label{tab:efusion_eval_results}
\centering
\begin{tabular}{|l|c|c|c|c|c|}
\hline
\textbf{Systems} & \# Epochs & \multicolumn{2}{c|}{\textbf{SASV-EER (\%)}} & \multicolumn{2}{c|}{\textbf{min a-DCF}} \\
\cline{3-6}
& & Dev & Eval & Dev & Eval \\
\hline
Baseline2 \cite{jung2022sasv2022spoofingawarespeaker} & N/A & 3.10 & 6.54 & - & - \\
\hline
eFusion & 1 & 2.20 & 5.80 & 0.0700 & 0.1500 \\
\hline
eFusion & 5 & 2.00 & 5.10 & 0.0550 & 0.1370 \\
\hline
eFusion & 10 & 2.00 & 6.30 & 0.0520 & 0.1685 \\
\hline
\end{tabular}
\end{table}

As shown in \autoref{tab:efusion_eval_results}, training for 5 epochs achieves a balance between underfitting and overfitting. Training for fewer epochs results in higher EER and min a-DCF values for both the development and evaluation datasets. Conversely, while training for more epochs lowers these metrics on the development set, it leads to increased EER and min a-DCF on the evaluation set, suggesting overfitting to the spoof attacks present in the development and training datasets. Furthermore, the eFusion model outperforms the baseline model, demonstrating that the introduced modifications to the model and the addition of VoxCeleb1 data enhance performance.

\subsection{ATMM-SAGA}
In this subsection, we present our experimental findings on the different variants of the ATMM-SAGA system described in~\autoref{subsection:ATMM_SAGA}. We begin by examining how different training approaches influence model performance and generalization to unseen attacks. Subsequently, we evaluate the various strategies for integrating the CM score.
\subsubsection{Comparison of Training Approaches}
We experimented with various training approaches while employing the S1 integration strategy. The experiments included adjustments to the training algorithm described in Alg.~\ref{alg:atmm} and the use of regularization methods, including \textit{dropout} and BN, to improve training stability and mitigate overfitting. Through a systematic evaluation of different training configurations, we aimed to determine the combination that best enhances model generalization.

\begin{table}[t]
\caption{Comparison of different training technique combinations under the S1 integration strategy, assessed using min a-DCF and SASV-EER on the ASVspoof2019 LA development and evaluation sets.}
\label{table:atmmsaga_train_tech}
\centering
\renewcommand{\arraystretch}{1.3}
\begin{tabular}{|c|c|c|c|c|c|c|}
\hline
\multicolumn{2}{|c|}{\textbf{min a-DCF}} & \multicolumn{2}{c|}{\textbf{EER (\%)}} & \textbf{BN} & \textbf{Drop} & \textbf{ATMM} \\
\cline{1-4}
\textbf{Dev} & \textbf{Eval} & \textbf{Dev} & \textbf{Eval} & & & \\ 
\hline
$\mathbf{0.0500}$ & 0.1464 & $\mathbf{1.46}$ & 5.74 & \xmark & \xmark & \xmark \\
\hline
0.1189 & 0.1386 & 4.45 & 5.22 & \cmark & \xmark & \xmark \\
\hline
0.0653 & 0.1422 & 2.25 & 5.58 & \xmark & \cmark & \xmark \\
\hline
0.1079 & 0.1315 & 4.48 & 4.98 & \cmark & \cmark & \xmark \\
\hline
0.0603 & $\mathbf{0.0480}$ & 2.31 & $\mathbf{2.18}$ & \xmark & \xmark & \cmark \\
\hline
0.0975 & 0.0702 & 6.60 & 4.21 & \cmark & \xmark & \cmark \\
\hline
0.0620 & 0.0516 & 2.42 & 2.27 & \xmark & \cmark & \cmark \\
\hline
0.0937 & 0.0707 & 6.84 & 4.17 & \cmark & \cmark & \cmark \\
\hline
\end{tabular}
\end{table}

The results in~\autoref{table:atmmsaga_train_tech} show that although conventional regularization techniques, such as BN and \textit{dropout}, are effective in certain cases, modifications to the training procedure, as described in Alg.~\ref{alg:atmm}, result in more substantial performance improvements. Notably, the results in~\autoref{table:atmmsaga_train_tech} show that applying ATMM without BN and without \textit{dropout} achieves the lowest min a-DCF and SASV-EER on the evaluation set. This suggests that ATMM itself provides sufficient regularization and that additional regularization may induce underfitting. At the same time, this configuration yields inferior performance on the development set compared to the model trained without ATMM. This discrepancy can be attributed to the development set sharing the same spoofing attacks as the training set, suggesting that models trained with ATMM are less susceptible to overfitting to the attacks that are in the training set, and improve the generalization to the unseen attacks. 

\subsubsection{Comparison of CM Score Integration Strategies within the ATMM-SAGA framework}
\begin{table}[t]
\caption{Comparison of CM score integration strategies within the ATMM-SAGA framework, evaluated in terms of min a-DCF and SASV-EER, on the ASVspoof2019 LA dataset’s development and evaluation sets.}
\label{tab:strategy_comp_atmm}
\centering
\renewcommand{\arraystretch}{1.3}  
\begin{tabular}{|l|c|c|c|c|}
\hline
\textbf{Systems} & \multicolumn{2}{c|}{\textbf{SASV-EER (\%)}} & \multicolumn{2}{c|}{\textbf{min a-DCF}} \\
\cline{2-5}
 & \textbf{Dev} & \textbf{Eval} & \textbf{Dev} & \textbf{Eval} \\
\hline
S1 & 2.31 & 2.18 & 0.0603 & 0.0480 \\
\hline
S2 & 2.28 & 2.19 & 0.0571 & 0.0501 \\
\hline
S3 & \textbf{1.80} & \textbf{2.00}  & \textbf{0.0555} & \textbf{0.0476}\\
\hline
SF & 3.87 & 5.45 & 0.1087 & 0.1245 \\
\hline

\end{tabular}
\end{table}

In this subsection, we examine the effectiveness of different CM score integration strategies within the SASV system. As shown in~\autoref{tab:experiment_results}, SAGA consistently outperforms score fusion, with the S1 strategy achieving slightly better performance than S2 on the evaluation dataset. This may indicate that early integration is more effective than late integration, although the observed differences may not be statistically significant. The full integration strategy (S3) achieves the best overall performance. Within this setting, empirical results indicate that employing the weighted multi-task objective defined in~\autoref{equation:total_loss}, with $\lambda = 0.1$ for the countermeasure term and $\lambda = 0.9$ for the ASV term, yields the best overall performance.

\subsection{ELEAT-SAGA}
In this section, we describe the experiments performed on the ATMM-SAGA system with the aim of improving its performance, which ultimately led to the proposed ELEAT-SAGA system described in~\autoref{subsection:ELEAT}. The investigated enhancements include the use of \textit{early feature} (EF) representations for improved CM score estimation and the introduction of a bypass mechanism through EAT.

\subsubsection{Integrating Early Features for Computing CM Scores}  
We explored multiple variations of the branching approach described earlier in \autoref{subsection:ELEAT} by extracting feature embeddings from early layers, denoted as ${x_1, x_2, x_3}$, as illustrated in \autoref{fig:eatfi}. Here, ${x_1}$ is derived from the activations of the $\mathit{FC}_1$ layer.

\begin{table}[t]
\centering
\caption{Performance comparison of ATMM-SAGA with early feature branching configurations for CM score calculation, evaluated in terms of SASV-EER (\%) and min a-DCF on the ASVspoof2019 development and evaluation sets.}
\label{tab:atmmif_saga_results}
\renewcommand{\arraystretch}{1.3}  
\begin{tabular}{|l|c|c|c|c|}
\hline
\textbf{Systems} & \multicolumn{2}{c|}{\textbf{SASV-EER (\%)}} & \multicolumn{2}{c|}{\textbf{min a-DCF}} \\
\cline{2-5}
& \textbf{Dev}  & \textbf{Eval} & \textbf{Dev} & \textbf{Eval} \\
\hline
ATMM-SAGA S3 & \textbf{1.80} & 2.00  & 0.0555 & 0.0476\\
\hline
ATMM-SAGA S3 (\( x_2, x_3 \)) & 2.35 & \textbf{1.71} & 0.0520 & \textbf{0.0399}\\
\hline
ATMM-SAGA S3 (\( x_1, x_2, x_3 \)) & 3.03 & 1.95  & \textbf{0.0483} & 0.0479  \\
\hline
\end{tabular}
\end{table}

The experimental results presented in \autoref{tab:atmmif_saga_results} compare different ATMM-SAGA configurations integrating early CM processing features. Incorporating features from \( x_1, x_2, x_3 \) yielded mixed results, improving EER but slightly increasing the min a-DCF on the evaluation set. Using only \( x_2, x_3 \) improved overall performance on the evaluation set, but resulted in a significant increase in EER and a decrease in min a-DCF on the development set. This indicates that ATMM-SAGA S3 (with \( x_2, x_3 \)) might be the superior approach, as min a-DCF is a more reliable metric for evaluating SASV performance.

\subsubsection{Introducing the bypass (EAT)}
We further explored enhancing ATMM by introducing a bypass mechanism, converting it into EAT, as outlined in Alg.~\ref{alg:eat}. The core idea behind EAT is to strategically bypass the SAGA operation when processing ASV data, preventing the model from learning unintended dependencies from out-of-domain data. The results of this integration are shown in \autoref{tab:eatifi_saga_results}, comparing different ATMM-SAGA configurations with and without EAT.
\begin{table}[t]
\centering
\caption{Performance comparison of ATMM-SAGA S3 configurations with and without EAT and EF integration, evaluated in terms of SASV-EER (\%), and min a-DCF on the ASVspoof2019 development and evaluation sets.}
\label{tab:eatifi_saga_results}
\renewcommand{\arraystretch}{1.3}  
\resizebox{0.49\textwidth}{!}{
\begin{tabular}{|c|c|c|c|c|c|c|}
\hline
 \multicolumn{2}{|c|}{\textbf{SASV-EER (\%)}} & \multicolumn{2}{c|}{\textbf{min a-DCF}} & \textbf{EAT} & \textbf{EF} & \textbf{ATMM} \\
\cline{1-4}
 \textbf{Dev} & \textbf{Eval} & \textbf{Dev} & \textbf{Eval} & & &  \\ 
\hline
{1.80} & 2.00  & 0.0555 & 0.0476 & \xmark & \xmark & \cmark \\
\hline
1.35 & 1.50 & 0.0366 & 0.0382 & \cmark & \xmark & \xmark \\
\hline
2.35 & 1.71 & 0.0520 & 0.0399 & \xmark & \cmark & \cmark \\
\hline
\textbf{1.08} & \textbf{1.22} & \textbf{0.0363} & \textbf{0.0303} & \cmark & \cmark & \xmark  \\
\hline
\end{tabular}
}
\end{table}

The results depicted in \autoref{tab:eatifi_saga_results} show that while incorporating early features and applying the bypass modification to ATMM (using EAT) individually improve performance on the evaluation set in terms of min a-DCF. Combining both approaches, as illustrated in \autoref{fig:eatfi}, leads to a significant performance gain for both the development and evaluation sets. This suggests that integrating early features for computing the CM score can substantially enhance performance. However, incorporating early features from out-of-domain data may hinder this improvement. The best results were achieved after 3 epochs of the EAT algorithm, as described in Alg.~\ref{alg:eat}. This demonstrates that EAT is a highly efficient method for training an integrated system.

\subsection{Embeddings Analysis}
In this subsection, we analyze the behavior of CM embeddings extracted from both the AASIST and ELEAT systems, as illustrated in \autoref{fig:eatfi}. Specifically, we examine the embeddings at the input to the $\mathit{FC}_4$ layer, as well as the SASV embeddings at the input to $\mathit{FC}_7$ within the same system. This analysis was conducted across the Train, Dev, and Eval sets of the ASVspoof2019 dataset. For visualization, we applied UMAP for dimensionality reduction, fitting it on the Train set and projecting the 2D results for all subsets.

\begin{figure}[t]
    \centering
    \begin{subfigure}[b]{0.158\textwidth}
        \includegraphics[width=\textwidth]{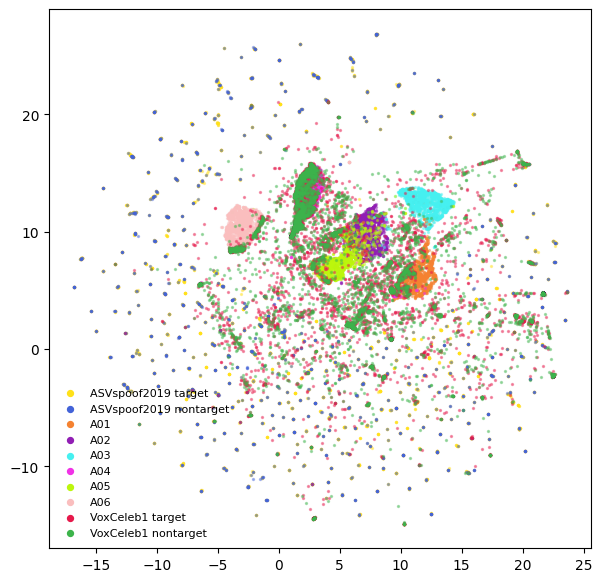}
        \caption{\footnotesize{Train AASIST}}
        \label{fig:emb_ast_trn}
    \end{subfigure}
    \begin{subfigure}[b]{0.158\textwidth}
        \includegraphics[width=\textwidth]{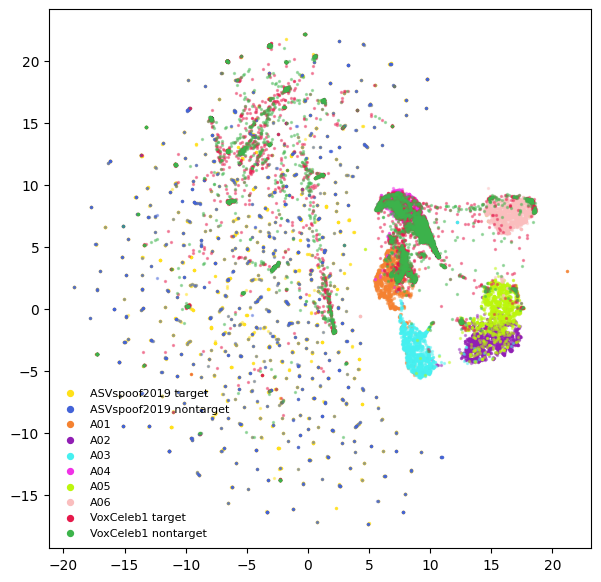}
        \caption{\footnotesize{Train ELEAT-CM}}
        \label{fig:emb_cm_trn}
    \end{subfigure}
    \begin{subfigure}[b]{0.158\textwidth}
        \includegraphics[width=\textwidth]{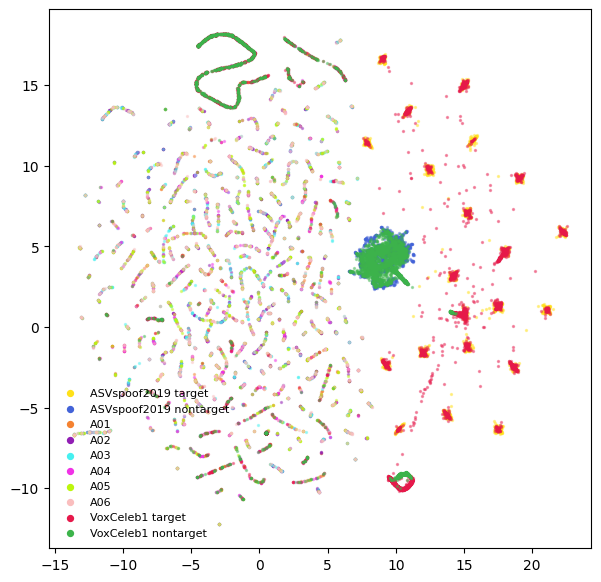}
        \caption{\footnotesize{Train ELEAT-SASV}}
        \label{fig:emb_sasv_trn}
    \end{subfigure}
    \begin{subfigure}[b]{0.158\textwidth}
        \includegraphics[width=\textwidth]{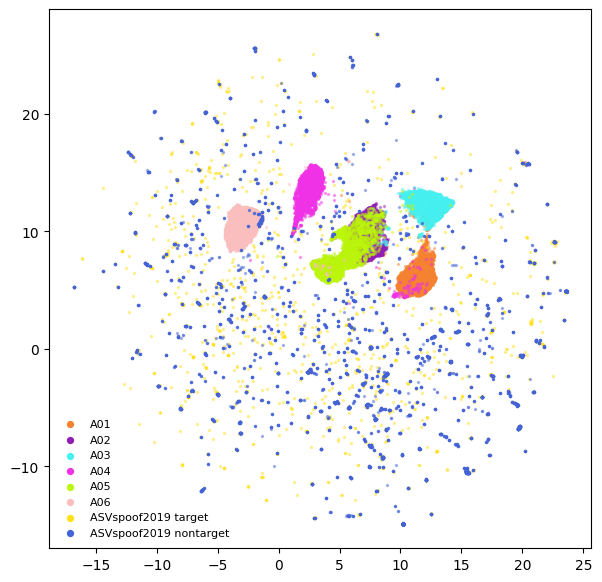}
        \caption{\footnotesize{Dev AASIST}}
        \label{fig:emb_ast_dev}
    \end{subfigure}
    \begin{subfigure}[b]{0.158\textwidth}
        \includegraphics[width=\textwidth]{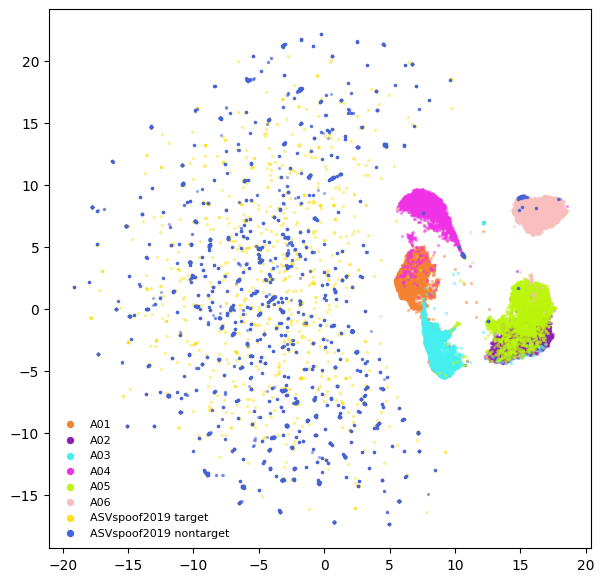}
        \caption{\footnotesize{Dev ELEAT-CM}}
        \label{fig:emb_cm_dev}
    \end{subfigure}
    \begin{subfigure}[b]{0.158\textwidth}
        \includegraphics[width=\textwidth]{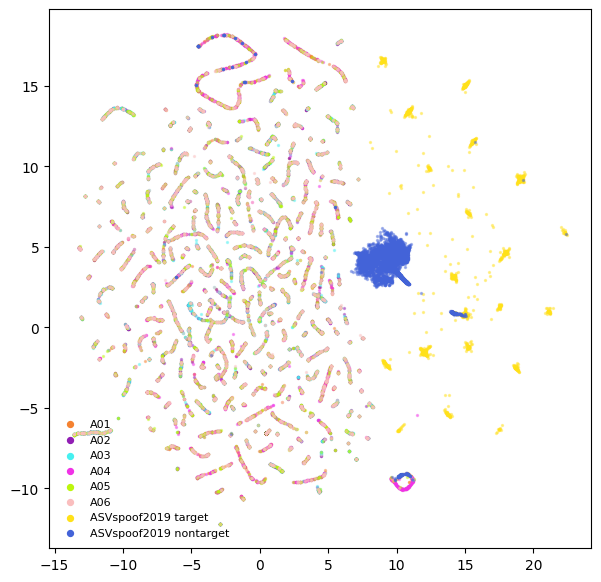}
        \caption{\footnotesize{Dev ELEAT-SASV}}
        \label{fig:emb_sasv_dev}
    \end{subfigure}
    \begin{subfigure}[b]{0.158\textwidth}
        \includegraphics[width=\textwidth]{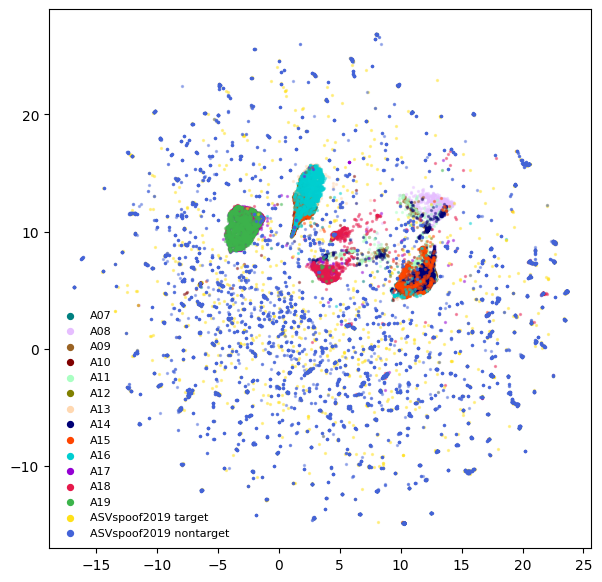}
        \caption{\footnotesize{Eval AASIST}}
        \label{fig:emb_ast_eval}
    \end{subfigure}
    \begin{subfigure}[b]{0.158\textwidth}
        \includegraphics[width=\textwidth]{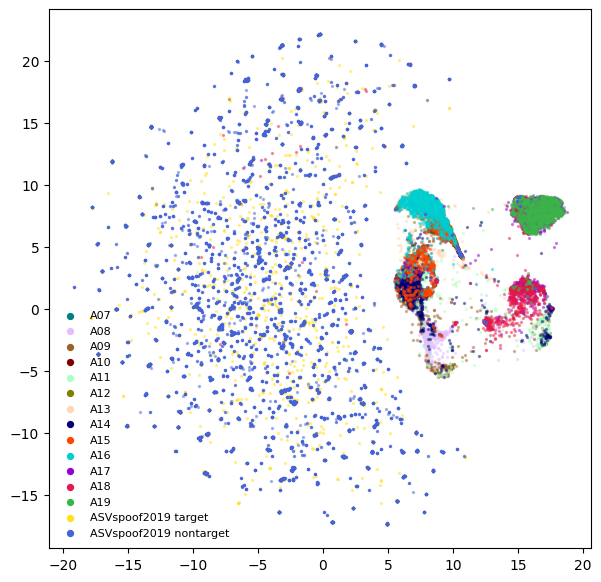}
        \caption{\footnotesize{Eval ELEAT-CM}}
        \label{fig:emb_cm_eval}
    \end{subfigure}
    \begin{subfigure}[b]{0.158\textwidth}
        \includegraphics[width=\textwidth]{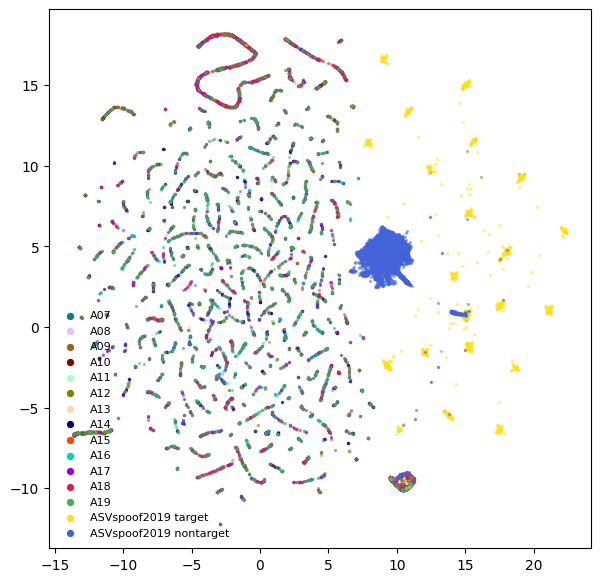}
        \caption{\footnotesize{Eval ELEAT-SASV}}
        \label{fig:emb_sasv_eval}
    \end{subfigure}
    
    \caption{UMAP visualization of embeddings from ASVspoof2019 Train, Dev, and Eval sets. Each row corresponds to one dataset split, while the columns show embeddings extracted from AASIST (left), ELEAT-CM (middle), and ELEAT-SASV (right).}
    \label{fig:umap_embeddings}
\end{figure}

As shown in~\autoref{fig:emb_ast_trn}~\autoref{fig:emb_ast_dev}~\autoref{fig:emb_ast_eval}, AASIST embeddings provide a clear separation between spoofed and bona fide samples, but show weaker separation for out-of-domain bona fide data such as VoxCeleb1, highlighting their limitations in generalization and motivating the need for the bypass mechanism in Alg.~\ref{alg:eat}. In contrast, ~\autoref{fig:emb_cm_trn}~\autoref{fig:emb_cm_dev}~\autoref{fig:emb_cm_eval} illustrate that ELEAT-CM embeddings display improved separability of spoof and bona fide data across all splits. The ELEAT-SASV embeddings, shown in ~\autoref{fig:emb_sasv_trn}~\autoref{fig:emb_sasv_dev}~\autoref{fig:emb_sasv_eval}, exhibit the most distinct class boundaries. Notably, SASV embeddings capture target, zero-effort non-target, and spoof classes with high separability, although they do not differentiate between spoofing attack types since spoof information is reduced to a single score. Interestingly, target embeddings form mini clusters that correspond to different speakers, whereas zero-effort non-target embeddings are clustered together.



\subsection{Results summary and baseline comparison}
\begin{table}[t]
\caption{Comparison of our systems with standalone systems, official SASV baselines, and state-of-the-art ECAPA-TDNN and AASIST based integration approaches, evaluated in terms of SASV-EER on the ASVspoof2019 LA development and evaluation sets.}
\label{tab:experiment_results}
\centering
\renewcommand{\arraystretch}{1.3}

\begin{tabular}{|l|c|c|c|}
\hline
\textbf{Systems} & \multicolumn{2}{c|}{\textbf{SASV-EER (\%)}} \\
 & Dev & Eval\\
\hline
ECAPA-TDNN \cite{Desplanques-2020} & 17.31 & 23.84 \\
AASIST \cite{jung2021aasistaudioantispoofingusing} & 15.86 & 24.38  \\
\hline
Liu et al. \cite{Liu-2024} & -- & 8.62 \\
Baseline2 \cite{jung2022sasv2022spoofingawarespeaker,shim22_odyssey} & 3.10 & 6.54  \\
eFusion & 2.00 & 5.10 \\
Mun et al. \cite{mun23_interspeech} & -- & 1.19\\
\hline
Baseline1-v2 \cite{jung2022sasv2022spoofingawarespeaker,shim22_odyssey} &1.01 & 1.71  \\
Zhang et al. \cite{probfusi} &1.10 &1.53  \\
Martín-Donas et al. \cite{vicomtech22_iberspeech} &0.58 &0.84  \\
\hline
ATMM-SAGA S3 & 1.80 & 2.00  \\
ELEAT-SAGA & 1.08 & 1.22  \\
\hline
\end{tabular}
\end{table}
As shown in Table~\ref{tab:experiment_results}, standalone systems such as ECAPA-TDNN~\cite{Desplanques-2020} and AASIST~\cite{jung2021aasistaudioantispoofingusing} perform poorly in the SASV setting, confirming that ASV and CM subsystems in isolation are insufficient. The Baseline2 system from the SASV challenge~\cite{jung2022sasv2022spoofingawarespeaker,shim22_odyssey} improves markedly through simple embedding-level fusion, while our modified eFusion model further enhances robustness. The Baseline1-v2 system~\cite{jung2022sasv2022spoofingawarespeaker,shim22_odyssey} achieves strong performance through calibrated score-level fusion. More sophisticated score fusion approaches~\cite{probfusi} further demonstrate the value of explicitly modeling cross-dependencies between ASV and CM information. Our proposed ATMM-SAGA S3 reduces gradient interference and overfitting to spoofing conditions, achieving 2.00\% SASV-EER on the evaluation set. Building on this, ELEAT-SAGA leverages early feature integration and task-aware bypassing, achieving SASV-EER of 1.22\%. As noted in~\cite{vicomtech22_iberspeech}, the authors also employ AASIST and ECAPA-TDNN to extract CM and ASV embeddings, respectively. However, they concatenate the test ASV and CM embeddings to compute a spoofing score that leverages ASV information within the CM pathway, linearly combining it with the ASV cosine similarity between enrollment and test embeddings. They found that including the enrollment embedding in the integration network degraded performance, thus it was excluded. The model, trained with a one-class softmax loss, achieved an SASV-EER of 0.84\%. Owing to the generality of our framework, these observations, together with the findings reported in~\cite{vicomtech22_iberspeech}, suggest a promising research direction. Specifically, future work may focus on deriving a stronger CM score and systematically exploring alternative loss functions to further enhance SASV performance.

\begin{table}[t]
\centering

\caption{Performance comparison of proposed SASV systems and official SASV baseline, evaluated in terms of min a-DCF and SASV-EER, on SpoofCeleb development and evaluation sets.}
\label{tab:eleat_saga_results_spoofceleb}
\resizebox{0.49\textwidth}{!}{
\renewcommand{\arraystretch}{1.3} 
\begin{tabular}{|l|c|c|c|c|c|}
\hline
\multirow{2}{*}{\textbf{Systems}} & \multirow{2}{*}{\textbf{Pretrained ASV}}
& \multicolumn{2}{c|}{\textbf{SASV-EER (\%)}} 
& \multicolumn{2}{c|}{\textbf{min a-DCF}} \\
\cline{3-6}
& & \textbf{Dev} & \textbf{Eval} & \textbf{Dev} & \textbf{Eval} \\
\hline
\multirow{2}{*}{Baseline \cite{spoofceleb}}
& ECAPA-TDNN  & -- & -- & -- & -- \\
& SKA-TDNN   & -- & -- & 0.3101 & 0.2902 \\
\hline
\multirow{2}{*}{SAGA S3}
& ECAPA-TDNN & 6.93 & 7.77 & 0.1272 & 0.1157 \\
& SKA-TDNN   & \textbf{4.79} & {5.05} & \textbf{0.0956} & 0.0812 \\
\hline
\multirow{2}{*}{ATMM-SAGA S3}
& ECAPA-TDNN & 7.43 & 8.22 & 0.1303 & {0.1082} \\
& SKA-TDNN   & 5.44 & {5.05} & 0.1123 & {0.0804} \\
\hline
\multirow{2}{*}{ELEAT-SAGA}
& ECAPA-TDNN & 7.23 & 7.90 & 0.1419 & 0.1151 \\
& SKA-TDNN    &{4.96} & \textbf{5.02} & 0.1088 & \textbf{0.0775}
\\
\hline
\end{tabular}
}
\end{table}
\subsection{ELEAT-SAGA on SpoofCeleb}
In the following, we evaluate the proposed systems on the SpoofCeleb dataset. The AASIST model is pretrained on the SpoofCeleb training partition and achieves EERs of 1.00\% and 0.26\% on the Dev and Eval sets, respectively. We reuse the same ECAPA-TDNN model employed in our previous experiments, pretrained on VoxCeleb1 and VoxCeleb2. On SpoofCeleb, this model yields EERs of 8.11\% on the Dev set and 10.57\% on the Eval set. To avoid any possibility of audio-level data leakage, we exclude VoxCeleb1 E and H from the SASV training pool. This precaution is necessary because SpoofCeleb is constructed from VoxCeleb1 recordings, and utterances from the E and H splits may appear in the SpoofCeleb evaluation set. We adjust the weighting parameter $\lambda$ in Alg.~\ref{alg:atmm} and Alg.~\ref{alg:eat} to accommodate this dataset. Empirically, the best performance is obtained with $\lambda = 0.035$ when $p=0$. We further compare our systems against the official SASV baseline released by the SpoofCeleb authors. The architecture of the official SASV baseline released by the SpoofCeleb authors~\cite{spoofceleb} is described in~\cite{mun23_interspeech}. Although this baseline also relies on an AASIST model pretrained on SpoofCeleb, we were unable to obtain the exact pretrained weights used in their implementation. Moreover, their system incorporates the SKA-TDNN model~\cite{mun2023frequency}, which is known to perform particularly well on this dataset. In our evaluation, the SKA-TDNN model achieves EERs of 2.76\% and 4.14\% on the Dev and Eval sets, respectively. In their SASV system~\cite{mun23_interspeech}, a single unified SASV encoder is trained using a three-stage procedure. This process consists of ASV pretraining on VoxCeleb2, subsequent adaptation using large-scale copy-synthesis spoof data with SASV-oriented loss functions, and a final fine-tuning stage on ASVspoof2019 LA to explicitly learn spoofing artifacts. As shown in \autoref{tab:eleat_saga_results_spoofceleb}, all of our proposed systems outperform the baseline. The performance differences among our systems are relatively small, which can be attributed to the fact that the speaker verification EER of the pretrained ASV systems is already relatively high. Consistent with the trends observed on ASVspoof2019, ATMM-SAGA yields lower evaluation min a-DCF than SAGA, while ELEAT-SAGA achieves the best overall performance among the proposed systems on SpoofCeleb. Beyond accuracy gains, ATMM substantially improves training efficiency, with models consistently converging in fewer than four epochs. Compared to the SAGA variant trained without ATMM, ATMM-SAGA converges approximately four times faster in terms of training epochs.

\section{Conclusion}
\label{sec:conclusion}
Our experiments show that small design choices strongly influence SASV performance. Leveraging early CM features, carefully avoiding BN/dropout, and bypassing CM training on noisy out-of-domain bona fide data improved generalization. These findings highlight that optimizing training procedures is as critical as architectural changes. We explored multiple integration strategies between speaker verification and countermeasure subsystems. We identified several key insights that inform the design of future SASV systems. Replacing CM embeddings with CM scores, as implemented in SAGA, simplifies the learning objective by removing the need to model dependencies among three high-dimensional embeddings. We further proposed ATMM, a training regime that alternates focus between speaker verification and spoofing countermeasure modules. Compared to basic joint training, ATMM exhibits improved generalization to unseen spoofing attacks and reduces overfitting to individual sub-tasks, and enables substantially more efficient training. Building upon ATMM, we introduced ELEAT, which leverages early CM features for score computation and selectively bypasses SAGA when training with out-of-domain speaker verification data. This strategy increases robustness and training efficiency, yielding strong performance within just a few training epochs. Our findings show that full SAGA integration consistently outperforms early, late, and score-level fusion approaches, confirming that multi-level spoof-awareness enhances discriminative capacity and overall system performance. We evaluated our best-performing model on both the ASVspoof 2019 and SpoofCeleb in-the-wild datasets, where it consistently outperformed the corresponding baseline systems. Collectively, supported by experiments on two different ASV systems and two different datasets, these contributions offer a practical, efficient, and high-performing framework for generalizable SASV.



\section{Acknowledgments}
This work is supported by the Israel Innovation Authority under project numbers 82457 and 82458.

\ifCLASSOPTIONcaptionsoff
  \newpage
\fi

\bibliographystyle{IEEEtran}
\bibliography{references}

\vspace{-35pt}
\begin{IEEEbiography}
[{\includegraphics[width=1in,height=1.25in,clip,keepaspectratio]{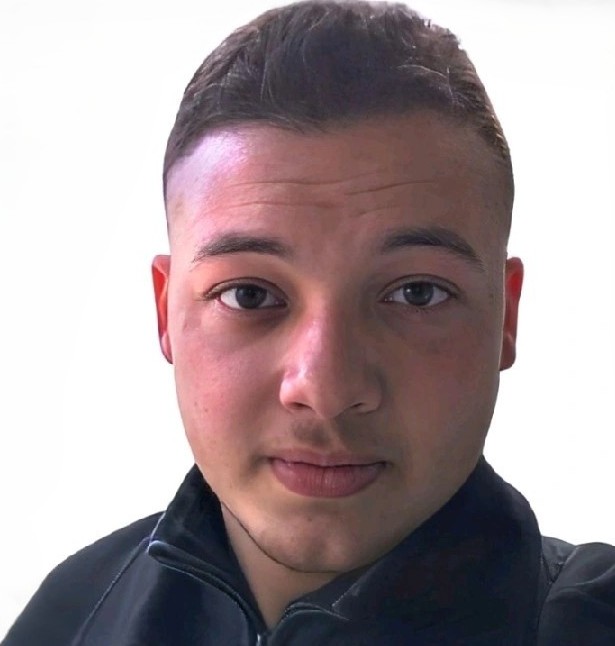}}]{Amro Asali}
Amro Asali received his B.Sc.\ in Computer Science from the Hebrew University of Jerusalem and his M.Sc.\ in Electrical and Computer Engineering from Ben-Gurion University of the Negev, where he is currently pursuing a Ph.D.\ in Electrical and Computer Engineering. His research interests include speaker verification and speech anti-spoofing.
\end{IEEEbiography}
\vspace{-35pt}
\begin{IEEEbiography}[{\includegraphics[width=1in,height=1.25in,clip,keepaspectratio]{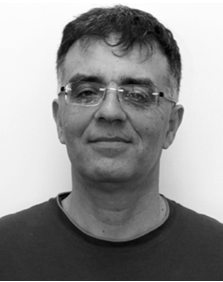}}]{Yehuda Ben-Shimol}
Yehuda Ben-Shimol (Member, IEEE) received the B.Sc., M.Sc., and Ph.D.\ degrees (Hons.) from Ben-Gurion University of the Negev, Beer-Sheva, Israel, all in electrical and computer engineering. He is currently a Senior Lecturer at the School of Electrical and Computer Engineering, Ben-Gurion University of the Negev. His main research areas include the design and analysis of communication networks, computer architecture, machine and deep learning (centralized and distributed), and neuromorphic computing.
\end{IEEEbiography}
\vspace{-35pt}
\begin{IEEEbiography}[{\includegraphics[width=1in,height=1.25in,clip,keepaspectratio]{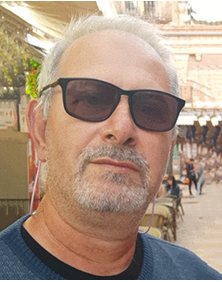}}]{Itshak Lapidot}
Itshak Lapidot received the B.Sc., M.Sc., and Ph.D.\ degrees from the Electrical and Computer Engineering Department at Ben-Gurion University of the Negev, Beer-Sheva, Israel. He held a postdoctoral position at the IDIAP Research Institute, Switzerland. He previously served as a Lecturer with the Electrical and Electronics Engineering Department at the Sami Shamoon College of Engineering and as a Researcher with the Laboratoire Informatique d’Avignon (LIA), University of Avignon, France. He is currently a Member of the Electrical Engineering Department at Afeka, Tel-Aviv Academic College of Engineering, Israel, and a Researcher with the Afeka Center of Language Processing. His research interests include speaker diarization, speaker clustering, speaker verification, anti-spoofing, and speech data assessment.
\end{IEEEbiography}

\end{document}